# Spectrum of the Dirac operator and inversion algorithms with dynamical staggered fermions [*]

T. Kalkreuter[a][†]

[a]Fachbereich Physik – COM, Humboldt-Universität, Invalidenstraße 110, D-10099 Berlin, Germany
Email: kalkreut@linde.physik.hu-berlin.de

Complete spectra of the staggered Dirac operator $\slashed{D}$ are determined in four-dimensional $SU(2)$ gauge fields with and without dynamical fermions. An attempt is made to relate the performance of multigrid and conjugate gradient algorithms for propagators with the distribution of the eigenvalues of $\slashed{D}$.

## 1. INTRODUCTION

Big efforts have been undertaken to find efficient multigrid (MG) methods for the computation of propagators in background gauge fields; see the lists of references in [1,2], and also [3]. Although ultimately one wants to simulate theories with dynamical fermions, all previous works focussed only on quenched gauge fields. However, it is reasonable to expect that MG methods have a chance to perform better when one considers "real" gauge fields which are generated in the presence of dynamical fermions. On the other hand one will not expect any big difference for the behavior of the conjugate gradient (CG) algorithm. The reasons for these two statements are as follows. The inclusion of the fermionic determinant in the Monte Carlo process will tend to decrease the number of (approximate) zero modes. This is so because configurations with less low-lying modes are more probable. MG methods intend to take care of the low-lying modes (which are responsible for critical slowing down) on coarser grids, and the task of dealing with a reduced number of low-lying modes should be easier. Concerning the CG algorithm, its (asymptotic) convergence properties are determined by the condition number. Since condition numbers of the (negative squared) massless staggered Dirac operator are not influenced dramatically by the presence of dynamical quarks, one does not expect a significant consequence for the convergence behavior of CG. In the present study we focus on the consequences of dynamical fermions for the performance of a variational MG algorithm which proved to work in the quenched case, but which was unable there to outperform CG [4].

## 2. MULTIGRID METHOD

The variational MG method for the solution of

$$(-\slashed{D}^2 + m^2)\chi = f \qquad (1)$$

is described in detail in Refs. [1,4]. We use a blocking procedure which is consistent with the symmetries of free staggered fermions. In this scheme coarsening is done with a factor of 3 [5]. Since the volumes of subsequent layers of the MG differ by a factor of 81, the actual implementaion of the MG method was only a twogrid algorithm. The kernel of the averaging operator is defined as the solution of a gauge covariant eigenvalue equation. In Ref. [5] our choice was given the name "Laplace choice", because we project on a block-local approximation of the "fermionic two-link Laplacian" $\Delta$. This $\Delta$ is defined through $\slashed{D}^2 = \Delta + \sigma_{\mu\nu} F_{\mu\nu}$, where $F_{\mu\nu}$ is the lattice definition of the field strength by means of plaquette terms.

## 3. SPECTRUM OF $-\slashed{D}^2$

As explained in the introduction, naively one expects $\slashed{D}$ to have less approximate zero modes in the presence of dynamical fermions than in the quenched case. In order to study this conjecture we need firstly a Hybrid Monte Carlo program, and secondly a method to determine the low-lying spectrum of $\not{D}$. For the generation of four-dimensional $SU(2)$ gauge fields coupled to dynamical staggered fermions a FORTRAN program with vectorized CRAY code was used, which had been written by S. Meyer and B. Pendleton. They used this program when they studied the chiral transition with many fermion flavors in the $SU(2)$ Higgs model [6]. Meyer's and Pendleton's program was used with four flavors of staggered fermions. The spectrum of $\not{D}$ was determined by means of a Lanczos procedure.

---
[*]to appear in the proceedings of LATTICE'94
[†]Work supported by Deutsche Forschungsgemeinschaft.

### 3.1. Lanczos procedure

The Lanczos method has been used in lattice field theory for a long time, see e. g. [7]. In the present exploratory study the complete spectrum of $-\not{D}^2$ was determined. This was done in order to be sure about the correctness (i. e. to have no numerical uncertainties) in the distribution of low-lying modes. Of the several computational variants of the Lanczos procedure the most stable one as described in [8, Algorithm 9.2.1 and remark on p. 492] was implemented. The "good" eigenvalues were determined by means of Cullum's and Willoughby's procedure [9]. More details can be found in Ref. [10].

### 3.2. Numerical results

One knows a priori that every eigenvalue of $-\not{D}^2$ in $SU(2)$ gauge fields is fourfold degenerate [11]. Thus, if there are no further degeneracies, then on a lattice $\Lambda$ of volume $|\Lambda|$ there must be $|\Lambda|/2$ different eigenvalues. Their sum must equal $\frac{1}{4}\mathrm{Tr}\,(-\not{D}^2) = 4|\Lambda|$. These statements are valid for periodic and for antiperiodic boundary conditions.

Concrete numerical investigations were done with the following parameters. The spectrum of $-\not{D}^2$ was investigated on $6^4$ and $12^4$ lattices. Two different kinds of boundary conditions (b.c.) were used. One with periodic b.c. in all directions for gauge and Fermi fields, and one with periodic b.c. for the gauge field in all directions and with antiperiodic b.c. on the Fermi field in time direction and periodic b.c. in spatial directions. The coupling $\beta = 4/g^2$ of the Wilson action for the $SU(2)$ gauge fields was varied between 1.8 and 5.0. Quark masses $m$ in the Hybrid Monte Carlo runs were chosen to be $m = 0.2$ and $m = 0.05$ (in units of $a = $ *twice* the spacing of the staggered lattice). These values were also used in Meyer's and Pendleton's work [6]; they quote $m = 0.1$ and $m = 0.025$ since they measured physical quantities in units of $a/2$.

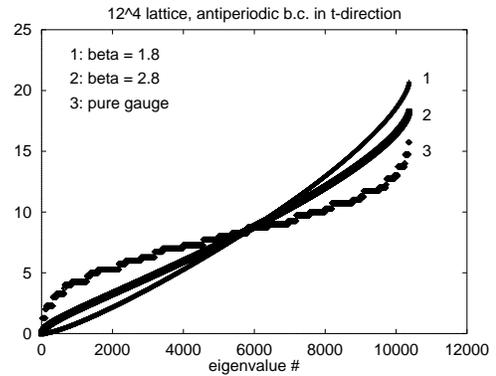

Figure 1. Spectrum of $-\not{D}^2$ on $12^4$ lattices with antiperiodic boundary conditions for the Fermi field in $t$-direction.

It turned out [10] that in nontrivial gauge fields there seem to be no additional degeneracies to the ones explained above. Only for $\beta = 5.0$ it was impossible to identify $|\Lambda|/2$ eigenvalues whose sum equals $4|\Lambda|$. For $\beta = 1.8,\ldots,2.8$ Cullum's and Willoughby's method works perfectly. On $6^4$ ($12^4$) lattices we always found 648 (10368) eigenvalues whose sum came out as $5184 + \delta_6$ ($82944 + \delta_{12}$), with $\delta_6 < 8\cdot 10^{-9}$ ($\delta_{12} < 1.7\cdot 10^{-6}$) in REAL arithmetic on a CRAY Y-MP. Because of the randomness of the nonvanishing off-diagonal matrix elements of $-\not{D}^2$ this is very good evidence that the spectrum was determined exactly.

Spectra on $12^4$ lattices are shown in Fig. 1. We number the different eigenvalues $\lambda_k$ by $k = 0, 1, 2, \ldots$, with $\lambda_0 < \lambda_1 < \ldots$ The data shown for finite $\beta$ are results obtained with a gauge field generated by the Hybrid Monte Algorithm in the presence of dynamical fermions with a mass of $m = 0.2$. However, on the scale of the whole spectrum there is very little difference compared to $m = 0.05$ or to a quenched gauge field. For finite $\beta$ there is also no difference between periodic and antiperiodic b.c. [1] The $\beta = \infty$ values are shown in Fig. 1 with their true multiplicities modulo the fourfold degeneracy mentioned above, so that the $\beta = \infty$ values indicate the curve which the numerical data should approach for large $\beta$.

When one looks at the complete spectra, figures for $6^4$ lattices look practically the same on the overall range as Fig. 1, when one rescales the abscissa by 16. (Only for the eigenvalues of the free $-\slashed{D}^2$ this is not true.)

Results for low-lying spectra and for condition numbers can be found in Ref. [10].

## 4. INVERSION OF $(-\slashed{D}^2 + m^2)$

A result of the present study is that the convergence behavior of CG in nontrivial gauge fields is practically only determined by the condition number $\kappa$ of $(-\slashed{D}^2 + m^2)$, and by the lattice size. This should be the case when the eigenvalues are distrubuted uniformely between the lowest and the highest one [12]. For configurations on a lattice of given size with the same $\kappa$, CG yields sequences of RMS norms of residuals which practically coincide, even if the spectra are different. As mentioned above, on the overall range of the spectra there is little difference between quenched simulations and simulations with dynamical fermions (of mass $m = 0.2, 0.05$). Therefore slight fluctuations in the distrubution of eigenvalues on small scales do not affect the convergence of CG. Thus if one wants to study the convergence of the CG algorithm one can do that with "cheap" quenched gauge fields, one does not have to take "expensive" unquenched configurations.

Results for the two-grid algorithm are as follows. An obvious statement is that convergence of the MG algorithm is not determined by $\kappa$. This is clear in the limiting case of free fields, because in pure gauges critical slowing down is completely eliminated by MG, i.e. convergence is completely independent of $\kappa$. In nontrivial gauge fields convergence of MG depends on details of the spectrum. Unfortunately, as in quenched gauge fields, inferiority of MG was found compared to CG, a factor of about 10 in CPU time. The poor performance of MG found earlier [4] is no feature of quenched computations. More details are in Ref. [10].

I wish to thank A. Brandt, A. Kennedy, G. Mack, S. Meyer, R. Sommer, and U. Wolff for discussions and/or comments. Financial support by Deutsche Forschungsgemeinschaft is gratefully acknowledged. (Present grant Wo 389/3-1.) The computations reported here were performed on the CRAY Y-MP of the Univ. of Kaiserslautern.

## REFERENCES

1. T.Kalkreuter, Nucl. Phys. B (Proc. Suppl.) 30 (1993) 257.
2. T. Kalkreuter, Nucl. Phys. B (Proc. Suppl.) 34 (1994) 768.
3. M. Bäker, preprint DESY 94-079 (May 1994), to appear in Int. J. Mod. Phys. C; these proceedings.
4. T. Kalkreuter, Ph. D. thesis and preprint DESY 92-158 (November 1992), shortened version to appear in Int. J. Mod. Phys. C5.
5. T. Kalkreuter, G. Mack, and M. Speh, Int. J. Mod. Phys. C3 (1992) 121.
6. S. Meyer and B. Pendleton, Phys. Lett. B241 (1990) 397; ib B253 (1991) 205.
7. I.M. Barbour, N.-E. Behilil, P.E. Gibbs, G. Schierholz, and M. Teper, in: The recursion method and its applications, Springer Series in Solid-State Sciences 58, eds. D. G. Pettifor and D. L. Weaire (Springer, Berlin, 1985).
8. G.H. Golub and C.F. v. Loan, *Matrix computations*, second edition, (The Johns Hopkins University Press, Baltimore, 1990).
9. J. Cullum and R.A. Willoughby, J. Comp. Phys. 44 (1981) 329.
10. T. Kalkreuter, preprint DESY 94-150, HUB-IEP-94/12 and KL-TH 19/94 (August 1994), to appear in Phys. Rev. D.
11. T. Kalkreuter, Phys. Rev. D48 (1993) 1926.
12. W. Hackbusch, *Iterative Lösung großer schwachbesetzter Gleichungssysteme* (B.G. Teubner, Stuttgart, 1991).

---

[1] This should be the case for expectation values of gauge invariant quantities when $-\mathbb{1}$ is an element of the gauge group. I wish to thank R. Sommer for this remark.